\documentclass[journal=ancac3,manuscript=article]{achemso}
\usepackage[version=3]{mhchem} 
\usepackage{siunitx}
\usepackage{amsmath,color}
\usepackage[Symbolsmallscale]{upgreek}

\SectionNumbersOn

\newcommand{\vect}[1]{\mathrm{\mathbf{{#1}}}}

\graphicspath{{figs/}}

\author{Sima Rek\v{s}tyt\.{e}}
\affiliation{Laser Research Center, Department of Quantum Electronics, Physics Faculty, Vilnius~University, Saul\.{e}tekio Ave. 10, LT-10223, Vilnius, Lithuania}
\author{Tomas Jonavi\v{c}ius}
\affiliation{Laser Research Center, Department of Quantum Electronics, Physics Faculty, Vilnius~University, Saul\.{e}tekio Ave. 10, LT-10223, Vilnius, Lithuania}
\author{Darius Gailevi\v{c}ius}
\affiliation{Laser Research Center, Department of Quantum Electronics, Physics Faculty, Vilnius~University, Saul\.{e}tekio Ave. 10, LT-10223, Vilnius, Lithuania}
\author{Mangirdas~Malinauskas}
\affiliation{Laser Research Center, Department of Quantum Electronics, Physics Faculty, Vilnius~University, Saul\.{e}tekio Ave. 10, LT-10223, Vilnius, Lithuania}
\email{mangirdas.malinauskas@ff.vu.lt}
\author{Vygantas~Mizeikis}
\affiliation{Research Institute of Electronics, Shizuoka University, 3-5-3-1 Johoku, Naka-ku, Hamamatsu 432-8561, Japan}
\author{Eugene~G.~Gamaly}
\affiliation{Research School of Physics and Engineering, Australian National University, Acton ACT 2601, Australia}
\author{Saulius~Juodkazis}
\affiliation{Center for Micro-Photonics, Faculty of Engineering and Industrial Sciences, Swinburne University of Technology, Hawthorn, VIC, 3122, Australia}
\alsoaffiliation{Melbourne Center for Nanofabrication, Australian National Fabrication Facility, Clayton, VIC 3168, Australia}
\email{sjuodkazis@swin.edu.au}

\title[An \textsf{achemso} demo]
{Nanoscale precision of 3D polymerisation via polarisation control}

\begin{document}
\newpage
\begin{abstract}
A systematic analysis of polarization effects in a direct write
femtosecond laser 3D lithography is presented. It is newly shown that
coupling between linear polarization of the writing light electric field
and temperature gradient  can be used to fine-tune feature sizes
in structuring of photoresists at a nanoscale.
The vectorial Debye focusing is used to simulate polarization
effects and a controlled variation up to $\sim$20\% in the
linewidth is shown experimentally for the identical axial extent of the
polymerised features. The revealed
mechanisms are relevant for a wide range of phenomena of
light-matter interaction at tight focusing in laser-tweezers and
in plasmonic or dielectric sub-wavelength focusing where strong
light intensity and thermal gradients coexist.
\end{abstract}



Direct laser writing (DLW) is widely used for three-dimensional
(3D) material processing and printing. One of its realizations via
polymerization enables free-form fabrication of
micro-/nano-objects down to tens-of-nanometers in spatial
resolution~\cite{KawataNat01,kim,fourkas,05nt846}. DLW in polymers
is based on curing a resist (resin) by nonlinear absorption at the
focal volume of a tightly-focused high peak power pulsed
laser radiation. Only a very small volume, which can be sub-100~nm
in cross sections, around the focal spot is affected during the
pulsed exposure~\cite{13lpr22,13nl3831,13pr1}. Pure optical light delivery
at tight focusing cannot explain the final size of the structure
due to the threshold effect of laser modification via
polymerization/cross-linking, thermal effects within the 3D focal
volume, and wet chemistry development which all affect the final
shape and size of the 3D
structure~\cite{09ol566,10lfz135,14lpr882,14ol3034,15lpr706}. Heat
accumulation at focus defined by pulse repetition rate and scan
speed is used to increase productivity of 3D polymerization and
makes thermal issues very important and actively
debated~\cite{11nl4218,12oe29890,13apl123107}. Polarization
effects in laser fabrication in 2D and 3D geometries are now
explored in polymerization by
DLW~\cite{Sun,03apl819,14ol6086,16acsn}. Also,
stimulated-emission-depletion (STED) control of 3D focal
volume~\cite{wegener,13oe10831}, orientation of deposition of
self-organized materials~\cite{Xu,16am}, melting and oxidation of thin
films~\cite{Oktem}, laser ablation~\cite{Simanovskii,Guay}, and as
self-organized nano-patterns induced on surface~\cite{14pqe119}
are among other polarisation related phenomena demonstrated
recently.

Here a systematic analysis via modeling and experiments is
presented to reveal polarization effects, their influence on the
feature size (resolution), and the coupling between thermal gradient and polarisation in DLW. 

\begin{figure}[tb!]
\centering
\includegraphics*[width=0.75\columnwidth]{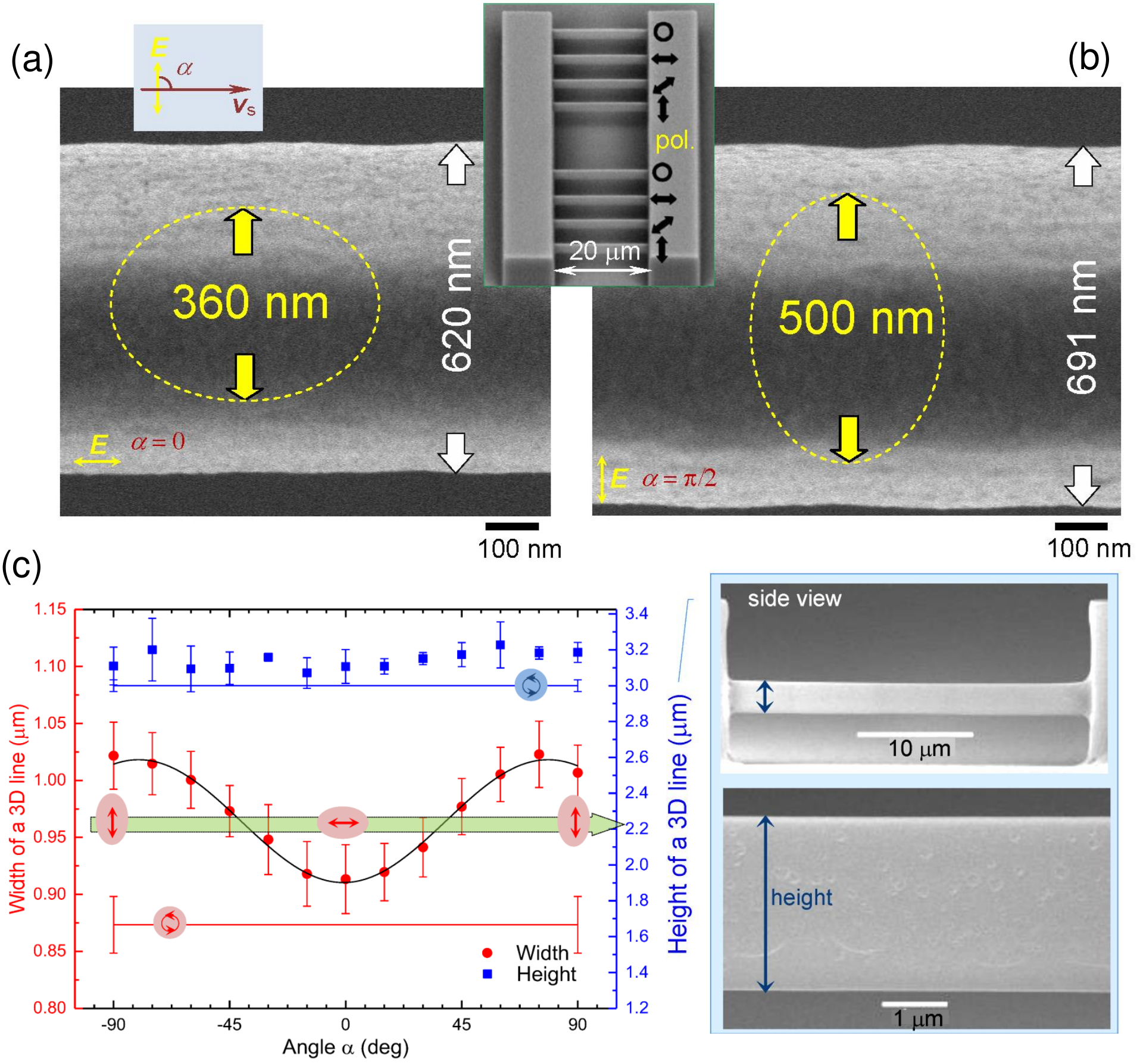}
\caption{(a-b) Top-view SEM images of 3D suspended bridges  formed
out of SZ2080 with 1\% BIS photo-initiator at the same conditions
(scanned along the x-axis at velocity $\vect{v}_s$) except for
light polarization direction:  $\vect{E}\parallel
\vect{v}_\mathrm{s}$ (or angle $\alpha = 0^\circ$) and
$\vect{E}\perp \vect{v}_\mathrm{s}$ (or angle $\alpha =
90^\circ$). The width difference is 10.3\% for $NA=1.4$ and
$\lambda = 1030$~nm. The ovals show the FWHM intensity cross
section calculated by vectorial Debye formula $W_\mathrm{l}\times
W_\mathrm{s} \equiv500\times 360$~nm (see Appxendix~A for
details). Note, the heat accumulation due to difference of number
of pulses per diameter is
$T_\mathrm{N}^{\mathrm{long}}/T_\mathrm{N}^{\mathrm{short}} =
1.59$. The inset shows a typical SEM image used for determination
of the height of the structures; $height/width$ from $\approx
3.07$ to $3.44$. (c) The width and height of 3D suspended bridges
at different angles, $\alpha$, between scan direction and linear
polarization. Numerical aperture $NA = 1.4$, pulse energy
$E_\mathrm{p} = 0.97$~nJ (at the focus) at 200~kHz repetition
rate. The fit line to width data is plotted as $\sim
\sin(\varphi)$; arrow marks direction of scan. } \label{f-wowther}
\end{figure}

To study and demonstrate the polarisation effects, the 3D suspended resolution bridges
at various angles, $\alpha$, between the linear polarisation and
scanning direction were fabricated on a glass substrate
(Fig.~\ref{f-wowther} inset in (a); see details in the
Experimental section).  The difference of the line widths was
10 - $20\%$ (varying exposure) at typical polymerization conditions for the linearly
polarized pulses. The largest width of a 3D
bridge was observed when scan direction was perpendicular to the
orientation of linear polarization, $\alpha = \pi/2$
($\vect{E}\perp \vect{v}_\mathrm{s}$). The height-to-width ratio
of the suspended lines was dependent on orientation of linear
polarization and was changing from $3.07$ to $3.44$, a
self-focusing was not present at our experimental
conditions.~\footnote{It was checked that the use of VIS-corrected
(400 - 700 nm) microscope objective lenses for the IR (1030 nm)
laser radiation was not causing additional spherical aberrations
and was not increasing the aspect ratio of polymerized bridge lines.
The aspect ratio at 515 nm wavelength was 3.00-3.57, almost the
same as for 1030 nm case (3.07-3.44) varying polarization
direction at intermediate exposure power/intensity. } Detailed
analysis of polarisation, threshold, and heat accumulation effects
which are all important are discussed next.

There are known effects of polarization on the scalar parameters
of laser-matter interaction, such as absorption coefficient and
ionization rate~\cite{Temnov,10prb054113,Gam,Rode}. It is also
known that heat conduction flux (vector) in plasma might be
depending on the direction of imposed field~\cite{Lifshitz}. In
what follows these effects are considered in succession:
\emph{(i)} the accumulation from the multiple pulses, \emph{(ii)}
effects of polarization under high-$NA$ focusing, and \emph{(iii)}
influence of the external high frequency electric field on the
electronic heat conduction. The later contribution has not yet
been considered in laser fabrication under tight focusing. All
these photomodification mechanisms occur simultaneously and affect
polymerization which takes approximately a millisecond for common
photoresists~\cite{14am6566} at a $\textgreater$ 90\% voxel
overlap at typical writing velocity of \SI{100}{\micro\meter/s} for widespread
laser 3D nanolithography~\cite{16lsa}.


For the used experimental conditions the focal spot diameter (at
$1/e^2$) can be calculated as $d_\mathrm{f} = 1.22\lambda/NA =
898$~nm assuming Gaussian intensity profile for simplicity.
However, at such tight focusing the vectorial Debye theory
(the specifics~\cite{Nasse10} can be found in Appendix~A) has to be used which predicts an ellipsoidal
focal spot with two lateral cross sections: $W_\mathrm{l} =
790$~nm and $W_\mathrm{s} = 572$~nm for long and short cross
sections, respectively (or 500 and 360~nm at FWHM) for the actual
experimental conditions and a $\sin$-apodisation function typical
for commercial objective lenses.

The dwell time of each pulse at the focal spot of diameter,
$d_\mathrm{f}$, equals to $t_{\mathrm{dw}} =
d_\mathrm{f}/v_{\mathrm{scan}}$. Thus the number of pulses per
spot at the repetition rate $R_{\mathrm{rep}} = 2\times
10^5$~pulses/s equals to $N_{\mathrm{spot}} =
t_{\mathrm{dw}}\times R_{\mathrm{rep}}$. Heat diffusion
coefficient for cold resist is similar to that in silica,
$D_{\mathrm{diff}} = 10^{-3}$~cm$^2$/s~\cite{Stuart}. Thus, the
cooling time of the heated volume, $t_{\mathrm{th}} =
d_\mathrm{f}^2/D$, has to be compared with the time gap between
subsequent pulses - \SI{5} {\micro\second}. The heat transfer to the
surrounding cold material between the pulses results in the
average temperature drop at the arrival of the next pulse and the
temperature accumulation can be explicitly calculated for the $N$
pulses as~\cite{Gamaly}:
\begin{equation}\label{e-ac}
T_\mathrm{N} = T_1(1 + \beta + \beta^2 + ... +\beta^N) =
T_1\frac{1-\beta^N}{1-\beta}\,,
\end{equation}
\noindent where
$\beta=\sqrt{\frac{t_{\mathrm{th}}}{t_{\mathrm{th}} +
1/R_{\mathrm{rep}}}}$\,. Assuming the same temperature jump, $T_1
= const$, for one pulse regardless polarization the accumulation
for the two orientations of polarization with respect to the scan
direction at fixed speed $v_\mathrm{s}$ =
\SI{100}{\micro\meter/\second} should result in significant
differences in heat accumulation: $T_\mathrm{N} \simeq 6.39 T_1$
($\alpha = 0$; Fig.~\ref{f-wowther}(b)) along $W_\mathrm{l}$
direction and $T_\mathrm{N} = 4.02T_1$ ($\alpha = \pi/2$ along
$W_\mathrm{s}$) due to different number of overlapping pulses $N$.
However, in the performed experiments the expected difference of
the line widths up to $\sim 59\%$ was not observed and was up to $20\%$ (applying highest, near optical damage intensities) whereas about 10\% was a typical deviation (applying intermediate intensities, commonly used for routine DLW lithography fabrication). This matches the previously reported findings that the shrinkage of polymerized features is conversely proportional to the applied exposure power (intensity)~\cite{09oe2143}. 

\begin{figure}[tb]
\centering
\includegraphics*[width=0.75\columnwidth]{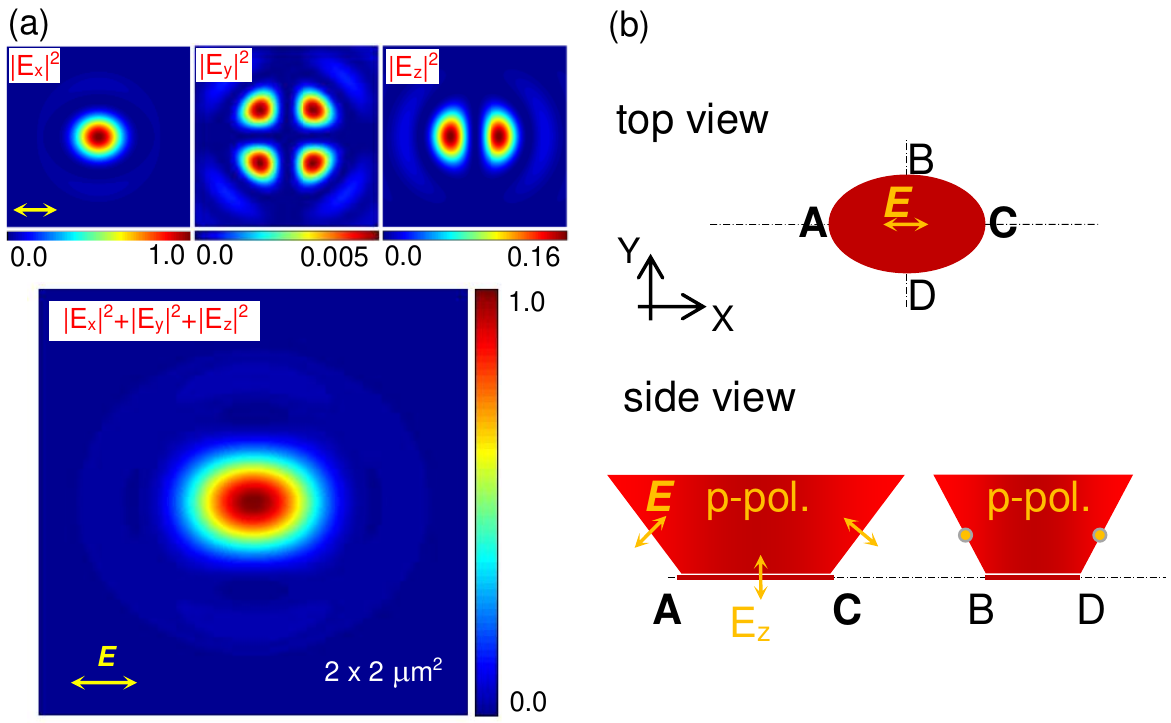}
\caption{(a) The calculated focal intensity distribution of a
linearly x-polarized 1030~nm wavelength beam under $NA = 1.4$
focusing using vectorial Debye theory with $\sin$-apodization
function (see Appendix A for details)~\cite{Chon}; normalized squares
of components of E-field (top row) and their sum
(bottom row). E-field polarization of the incident light is shown
by arrows. (b) Schematic view of polarization under tight focusing, which acquires p- and s-polarization and longitudinal
$E_\mathrm{z}$ components.} \label{f-lin}
\end{figure}

It was found theoretically that the polarization state affects the
shape of the focal spot when the laser beam focuses with high-NA
lenses~\cite{Shamir,Sun}. Effect starts at $NA = 0.2$ and at 0.95
it is significant however not quantified~\cite{Shamir}. The energy
density distribution, $w_{\mathrm{en}} =
(\vect{E}\cdot\vect{E}^*)/16\uppi $ of linearly polarized beam has
an elliptical form with the long axis in the E-field direction
(Fig.~\ref{f-lin}(a)). Besides in the central part of the tightly
focused beam the polarization state becomes unidentifiable.

The high-$NA$ focusing and polarization around the focal spot of
p-polarized (parallel to the plane of incidence) beam are
considered next (Fig.~\ref{f-lin}(b)). One can see that in the
direction $A-C$ (plane of incidence) across the focus polarization
changed projection with increased contribution of a perpendicular
to the sample surface component and is different from the
projection in the $B-D$ cross section. This effect significantly
increases for dense plasma for large angle of
incidence~\cite{Gibbon}. The p-pol. always has a stronger
absorption in a resist at pre-breakdown conditions while the beam
of circular polarization should experience an average absorption
of that between s- and p-polarizations (See Appendix~B for
details).

The difference in the energy absorption along the polarization
direction $\parallel$ might be significantly higher than in the
perpendicular direction $\perp$ resulting in elongation of the
absorbing area in the direction of the field as it was observed in
the experiments. It is noteworthy that due to the presence of both
lateral $E_\mathrm{x}$ and $E_\mathrm{y}$ fields
(Fig.~\ref{f-lin}(a)) the p-pol. will be always present at the
focal plane (Appendix~B). 

Analysis presented above shows how to account for polarisation
defined absorption which, in turn, defines local temperature in
the focal region. Following a general description, the heat
conduction flux in plasma placed into external electro-magnetic
field has a form~\cite{Lifshitz}:
\begin{equation}\label{e1}
q_\alpha = -\kappa_{\alpha\beta}\frac{\partial T}{\partial
x_\beta} = \kappa_1\frac{\partial T}{\partial x_\alpha} + \kappa_2
\mathbf{e}_\alpha \mathbf{e}_\beta \frac{\partial T}{\partial
x_\beta}.
\end{equation}
The first term is the conventional heat flow, while the second
term relates to the contribution due to the effect of external
field. Here, $\mathbf{e} = \vect{E}/E$  is the unit vector in the
direction of the electric field; $\kappa_{1,2}$ are two scalar
functions, which can be obtained from the solution of the kinetic
equation. Note that the general form of Eqn.~2 is
applicable to the constant external field as well as to the high
frequency field. The second term contribution holds after
averaging over the period of the light oscillation. Thus, it is
applicable to the experiments on polymerization by DLW. It follows
from Eqn.~2 that the field-related second term equals to
zero when the polarization direction is $\perp$ to the temperature
gradient, i.e. $\mathbf{e}_x \mathbf{e}_y \equiv 0$. Let us now
consider effects of linear and circular polarizations in
laser-plasma interactions.

\textbf{Stationary case.} Movement of the laser spot creates
different temperature gradients $\nabla T$ along and across the
scan direction $\vect{v}_\mathrm{s}$. It is instructive to
consider the most simple case of uniform heated spot and coupling
of heat transport with polarization as follows from Eqn.~2
even without scanning.

\emph{Linear polarization.} The intense laser beam along the
normal (z coordinate) produces plasma spot elongated in direction
of polarization (x,y plane) at focus, surrounded by a cold
unperturbed resist. Let us suggest that the temperature is
homogeneous in the interior of the focus. Therefore the
temperature gradients are concentrated at the outer boundary of
heated spot directed along the radii from the center to the
outside; polarization  is linear with the $\mathbf{E}$-field
direction along x-axis. As it follows from Eqn.~2 the field
affects the heat flow only along the x-axis, while in the other
parts of heated circle the heat flow is unperturbed because
temperature gradient and the field are perpendicular to each
other. Therefore expected effect of the linear polarization is an
elongation of the laser-affected spot in the polarization
direction turning a circle into an oval.

\emph{Circular polarization.} The electric field vector of the
incident beam at any moment of time is collinear to the
temperature gradient directed along the radius of the beam. Thus,
the field influence on a heat flow is evenly distributed along the
circle embracing the heated zone leading to the small change of
the zone's radius at the same absorbed energy density.


The effects of the high-$NA$ focusing and E-field enhanced heat
conduction both lead to elongation of the focal spot in the
direction of the E-field in the case of linear polarization.
Semi-quantitative estimate for elongation of polymerized region
could be calculated as the following. Conversion of s-pol. light
to the p-pol. along the polarization direction as described above
results in significant (up to $20\%$; see Appendix~B) increase in
absorption  and therefore increasing the temperature gradient at
the edges of the spot and promotes polymerization (cross-linking).
Simultaneously, heat accumulation is also present due to the
difference in the number of pulses per spot along the scan. The
discussed optical and thermal mechanisms are efficient during the
pulse time only. The elongation of the polymerized region might be
estimated as the distance which the heat wave travels during the
pulse time transferring the energy to the area unaffected by laser
directly, $L_{\mathrm{heat}} = \sqrt{D\tau} \simeq 5.5$~nm per
pulse. The effect from many consecutive pulses then accumulates in
a qualitative agreement with observations. It follows that the
additional polymerization length directly depends on the laser
fluence, wavelength, pulse duration and repetition rate: $\Delta
W\propto\tau\sqrt{I\lambda}$ (Appendix.~C).

Polarisation effects are ubiquitous among earlier results on
ablation~\cite{Guay}, nano-ripple formation (including using
relatively low $NA =
0.1-0.25$)~\cite{05prb125429,07ass4740,15ass}, growth of
nano-flakes in solution~\cite{Xu}, and controlled melting of
films~\cite{Oktem}. Elongation of fabricated features beyond extent of the focal spot is clearly shown using short laser pulses.
Even at low-$NA = 0.25$ focusing and ablation  of MgF$_2$ at very
high intensity~\cite{Simanovskii} there is clear elongation in
direction of linear polarisation which is comparable in size with
ablation pits for the $\tau = 1$~ps duration and $\lambda$ =
\SI{6.25}{\micro\meter} wavelength pulses.

\begin{figure}[tb]
\centering
\includegraphics*[width=0.75\columnwidth]{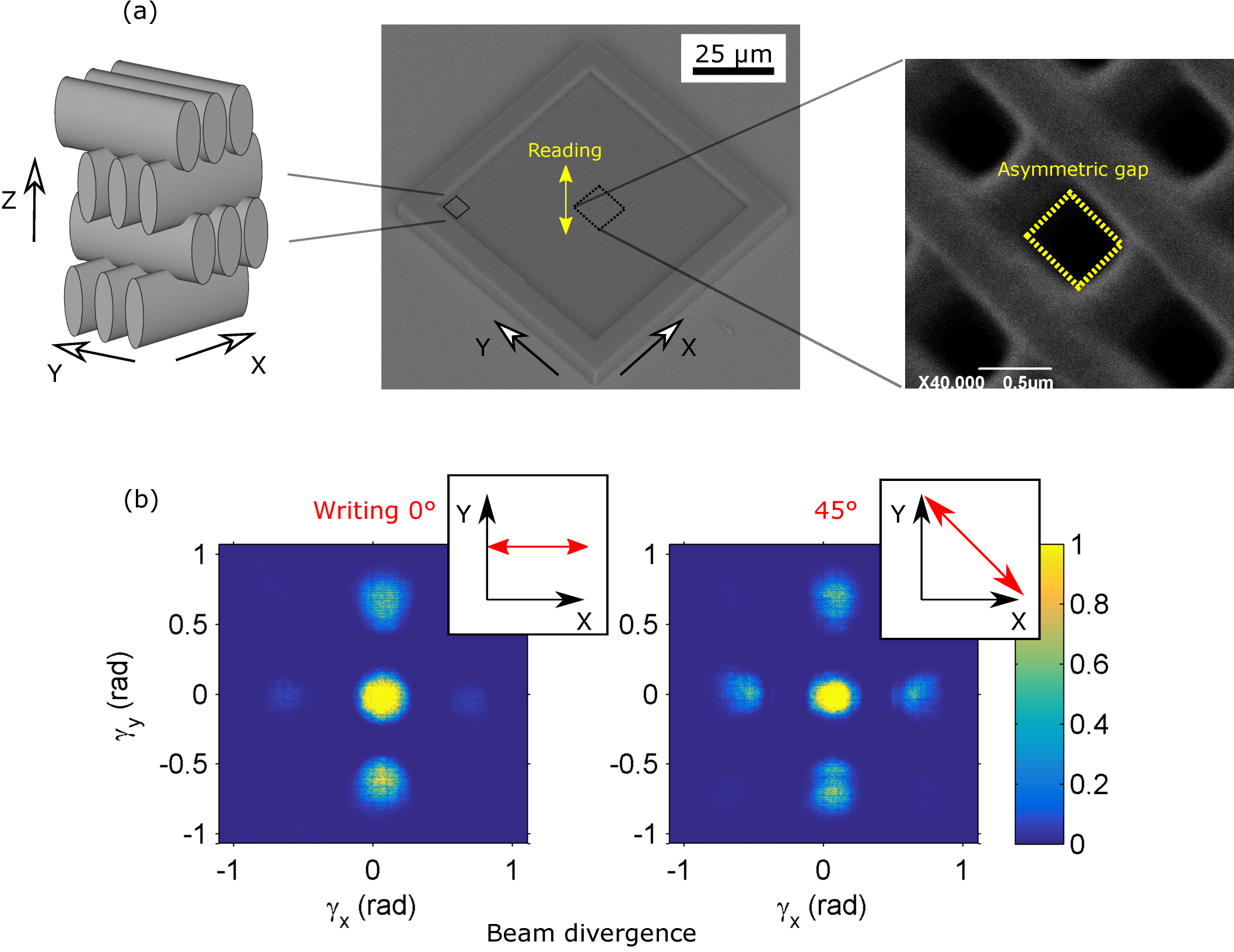}
\caption{(a) SEM image of a woodpile PhC fabricated with two
spatial periods, $dxy$ = 1 and $dz$ = 6 $\mu$m (2 full periods, 8 layers);
for a similar arrangement, see ref.~\cite{13ol2376}. Galvanometric
scanner was used for PhC polymerization and a monolithic wall was
made to reduce deformations. Writing parameters were: the pulse
energy $E_p$ = 0.73 nJ (corresponding to $I_p$ = 1 TW/cm$^2$
intensity. The line width difference in respect to each other due to polarization is 8\% resulting in an asymmetric gap shown in top right. (b)  shows diffraction pattern obtained by focusing
(NA = 0.25) a Gaussian He-Ne probe ($\lambda$ = 633 nm) on the PhC
and projecting the resulting pattern on a screen (a reference grid
artifact can be seen). An asymmetrical diffraction pattern for the
DLW polarization oriented along the PhC X-axis is clearly
recognizable on the left side whereby it is symmetrical if pre-compensated by writing direction on the right side.}
\label{f-phc}
\end{figure}

By controlling light energy delivery and absorption via
polarisation control described above, new opportunities to
exercise nanoscale precision in 3D polymerisation/printing are
opened. Next, a photonic crystal (PhC) structure was
polymerisied for the angular filtering and collimation of
light~\cite{13ol2376}. For such filters, polarisation induced
linewidth non-homogeneity causes spatially asymmetric performance
of PhC structures quantifiable by angular divergence in x-
and y-directions, $\gamma_{x,y}$. 
Figure~\ref{f-phc} shows the PhC structure and its performance as
collimating filter. Asymmetry in diffraction pattern (b) is a
signature of the $\sim 10\%$ line width difference (which is enough for the clear observation) when x-
linearly polarised laser beam was scanned along x- and
y-directions when compared with $\pi/4$-tilted beam which makes
the same angle with the scan direction.

The proposed analysis can qualitatively explain the action of linear polarized light on the width and symmetry of
the 3D suspended polymerized bridges and PhCs under tight
focusing. It is shown that polarization of incident laser field
allows controlling the polymerization process on a scale of
nanometers on already sub-wavelength 3D patterns.

Heat diffusion is reduced along the linear polarization and causes
local over-heating but does not affect heat flow in the
perpendicular direction during the pulse. Absorption is larger for
the p-pol. which contributes to elongation of the polymerized
features. 
The observed effect should be more pronounced for longer
wavelengths, higher pulse energies, and longer pulses according to
$\Delta W\propto\tau\sqrt{I\lambda}$. 
The coupling between the $\mathbf{E}$ and $\nabla T$ is
explained for the first time and provides new dimension to control
optically nanoscale processes. The ascribed findings demonstrate that the beam's polarization can be practically used as the voxel's aspect ratio tuning parameter and should be taken into account for precise 3D fabrication of nanostructures such as PhCs and micro-optical components, yet perhaps can be disregarded for some bulkier objects or larger scale scaffolds. 

\small\section*{Experimental}

\small{SZ2080 organic-inorganic hybrid material was doped with
4,4'-bis(diethylamino)-benzophenone (BIS) photoinitiator at 1\% by
weight and was used as a photopolymer
(IESL-FORTH)~\cite{08acsn2257}. A drop of photosensitive material
was heated for 20~min at each of 40, 70, and 90$^\circ$C
temperatures as a prebake prior to laser polymerization.  A
femtosecond Yb:KGW laser amplifier (Pharos, Light Conversion Ltd.)
radiation was used  at a central wavelength $\lambda$ = 1030~nm
with pulse repetition rate $R_{\mathrm{rep}}$ = 200~kHz and pulse
duration $\tau$ = 300~fs. We used middle range pulse energies from
an empirically established fabrication window when 3D suspended
bridges without structural damage are retrieved. This corresponded
to the 1.75~mW of average optical power (measured before the
objective) resulting in the 1.16~TW/cm$^2$
intensity/irradiance at the focus of a $100^\times$ magnification
objective lens of numerical aperture $NA = 1.4$; transmission of
the optical elements was taken into account. The diameter of the
collimated beam coming to the objective was of 8.5~mm and the
entrance aperture was 6 mm, thus an Airy disk at the focus in the
sample could be observed.}

\small{During fabrication the samples were translated at a
\SI{100}{\micro\meter/\second} linear velocity commonly used in
DLW 3D nanolithography~\cite{KawataNat01,kim,fourkas,08acsn2257,13nl3831,14am6566,16lsa}. After fabrication samples were developed
for 30~min in 4-methyl-2-pentanone bath and critical point drier was employed to avoid
structural changes due to capillary forces. All of the samples
were coated with 20~nm thick layer of gold using sputter coater
(note, the actual thickness of Au on the 3D suspended bridges was
smaller). Width and height of 3D suspended lines were examined
using scanning electron microscopy (SEM).}


\small
\begin{acknowledgement}

NATO SPS-985048 \emph{"Nanostructures for Highly Efficient Infrared Detection"} grant is acknowledged.

\end{acknowledgement}

\small\bibliography{lpr,references}

\providecommand{\latin}[1]{#1}
\providecommand*\mcitethebibliography{\thebibliography}
\csname @ifundefined\endcsname{endmcitethebibliography}
  {\let\endmcitethebibliography\endthebibliography}{}
\begin{mcitethebibliography}{47}
\providecommand*\natexlab[1]{#1}
\providecommand*\mciteSetBstSublistMode[1]{}
\providecommand*\mciteSetBstMaxWidthForm[2]{}
\providecommand*\mciteBstWouldAddEndPuncttrue
  {\def\EndOfBibitem{\unskip.}}
\providecommand*\mciteBstWouldAddEndPunctfalse
  {\let\EndOfBibitem\relax}
\providecommand*\mciteSetBstMidEndSepPunct[3]{}
\providecommand*\mciteSetBstSublistLabelBeginEnd[3]{}
\providecommand*\EndOfBibitem{}
\mciteSetBstSublistMode{f}
\mciteSetBstMaxWidthForm{subitem}{(\alph{mcitesubitemcount})}
\mciteSetBstSublistLabelBeginEnd
  {\mcitemaxwidthsubitemform\space}
  {\relax}
  {\relax}

\bibitem[Kawata \latin{et~al.}(2001)Kawata, Sun, Tanaka, and
  Takada]{KawataNat01}
Kawata,~S.; Sun,~H.-B.; Tanaka,~T.; Takada,~K. Finer features for functional
  microdevices. \emph{Nature} \textbf{2001}, \emph{412}, 697 -- 698\relax
\mciteBstWouldAddEndPuncttrue
\mciteSetBstMidEndSepPunct{\mcitedefaultmidpunct}
{\mcitedefaultendpunct}{\mcitedefaultseppunct}\relax
\EndOfBibitem
\bibitem[Lee \latin{et~al.}(2006)Lee, Yang, Park, and Kim]{kim}
Lee,~K.; Yang,~D.-Y.; Park,~S.~H.; Kim,~R.~H. Recent developments in the use of
  two-photon polymerization in precise {2D} and {3D} microfabrications.
  \emph{Polym. Adv. Technol.} \textbf{2006}, \emph{17}, 72--82\relax
\mciteBstWouldAddEndPuncttrue
\mciteSetBstMidEndSepPunct{\mcitedefaultmidpunct}
{\mcitedefaultendpunct}{\mcitedefaultseppunct}\relax
\EndOfBibitem
\bibitem[Maruo and Fourkas(2008)Maruo, and Fourkas]{fourkas}
Maruo,~S.; Fourkas,~J. Recent progress in multiphoton microfabrication.
  \emph{Laser Photon. Rev.} \textbf{2008}, \emph{2}, 100--111\relax
\mciteBstWouldAddEndPuncttrue
\mciteSetBstMidEndSepPunct{\mcitedefaultmidpunct}
{\mcitedefaultendpunct}{\mcitedefaultseppunct}\relax
\EndOfBibitem
\bibitem[Juodkazis \latin{et~al.}(2005)Juodkazis, Mizeikis, Seet, Miwa, and
  Misawa]{05nt846}
Juodkazis,~S.; Mizeikis,~V.; Seet,~K.~K.; Miwa,~M.; Misawa,~H. Two-photon
  lithography of nanorods in \textrm{SU-8} photoresist. \emph{Nanotechnol.}
  \textbf{2005}, \emph{16}, 846 -- 849\relax
\mciteBstWouldAddEndPuncttrue
\mciteSetBstMidEndSepPunct{\mcitedefaultmidpunct}
{\mcitedefaultendpunct}{\mcitedefaultseppunct}\relax
\EndOfBibitem
\bibitem[Fischer and Wegener(2013)Fischer, and Wegener]{13lpr22}
Fischer,~J.; Wegener,~M. Three-dimensional optical laser lithography beyond the
  diffraction limit. \emph{Laser Photon. Rev.} \textbf{2013}, \emph{7},
  22--44\relax
\mciteBstWouldAddEndPuncttrue
\mciteSetBstMidEndSepPunct{\mcitedefaultmidpunct}
{\mcitedefaultendpunct}{\mcitedefaultseppunct}\relax
\EndOfBibitem
\bibitem[Kabouraki \latin{et~al.}(2013)Kabouraki, Giakoumaki, Danilevicius,
  Gray, Vamvakaki, and Farsari]{13nl3831}
Kabouraki,~E.; Giakoumaki,~A.; Danilevicius,~P.; Gray,~D.; Vamvakaki,~M.;
  Farsari,~M. Redox Multiphoton Polymerization for 3D Nanofabrication.
  \emph{Nano Lett.} \textbf{2013}, \emph{13}, 3831--3835\relax
\mciteBstWouldAddEndPuncttrue
\mciteSetBstMidEndSepPunct{\mcitedefaultmidpunct}
{\mcitedefaultendpunct}{\mcitedefaultseppunct}\relax
\EndOfBibitem
\bibitem[Malinauskas \latin{et~al.}(2013)Malinauskas, Farsari, Piskarskas, and
  Juodkazis]{13pr1}
Malinauskas,~M.; Farsari,~M.; Piskarskas,~A.; Juodkazis,~S. Ultrafast-laser
  micro/nano-structuring of photo-polymers: a decade of advances. \emph{Phys.
  Rep.} \textbf{2013}, \emph{533}, 1--31\relax
\mciteBstWouldAddEndPuncttrue
\mciteSetBstMidEndSepPunct{\mcitedefaultmidpunct}
{\mcitedefaultendpunct}{\mcitedefaultseppunct}\relax
\EndOfBibitem
\bibitem[Takada \latin{et~al.}(2009)Takada, Wu, Chen, Shoji, Xia, Kawata, and
  Sun]{09ol566}
Takada,~K.; Wu,~D.; Chen,~Q.-D.; Shoji,~S.; Xia,~H.; Kawata,~S.; Sun,~H.-B.
  Size-dependent behaviors of femtosecond laser-prototyped polymer
  micronanowires. \emph{Opt. Lett.} \textbf{2009}, \emph{34}, 566--568\relax
\mciteBstWouldAddEndPuncttrue
\mciteSetBstMidEndSepPunct{\mcitedefaultmidpunct}
{\mcitedefaultendpunct}{\mcitedefaultseppunct}\relax
\EndOfBibitem
\bibitem[Malinauskas \latin{et~al.}(2010)Malinauskas, Bi{\v{c}}kauskait{\.{e}},
  Rutkauskas, Paipulas, Purlys, and Gadonas]{10lfz135}
Malinauskas,~M.; Bi{\v{c}}kauskait{\.{e}},~G.; Rutkauskas,~M.; Paipulas,~D.;
  Purlys,~V.; Gadonas,~R. Self-polymerization of nano-fibres and nano-membranes
  induced by two-photon absorption. \emph{Lith. J. Phys.} \textbf{2010},
  \emph{50}, 135--140\relax
\mciteBstWouldAddEndPuncttrue
\mciteSetBstMidEndSepPunct{\mcitedefaultmidpunct}
{\mcitedefaultendpunct}{\mcitedefaultseppunct}\relax
\EndOfBibitem
\bibitem[Liu \latin{et~al.}(2014)Liu, Sun, Dong, Yang, Chen, and Sun]{14lpr882}
Liu,~D.-X.; Sun,~Y.-L.; Dong,~W.; Yang,~R.-Z.; Chen,~Q.-D.; Sun,~H.-B. Dynamic
  laser prototyping for biomimetic nanofabrication. \emph{Laser Photon. Rev.}
  \textbf{2014}, \emph{8}, 882--888\relax
\mciteBstWouldAddEndPuncttrue
\mciteSetBstMidEndSepPunct{\mcitedefaultmidpunct}
{\mcitedefaultendpunct}{\mcitedefaultseppunct}\relax
\EndOfBibitem
\bibitem[Jiang \latin{et~al.}(2014)Jiang, Zhou, W.X., Gao, Huang, Jiang,
  Baldacchini, Silvain, and Lu]{14ol3034}
Jiang,~L.; Zhou,~Y.; W.X.,; Gao,~Y.; Huang,~X.; Jiang,~L.; Baldacchini,~T.;
  Silvain,~J.-F.; Lu,~Y. Two-photon polymerization: investigation of chemical
  and mechanical properties of resins using Raman microspectroscopy. \emph{Opt.
  Lett.} \textbf{2014}, \emph{39}, 3034--3037\relax
\mciteBstWouldAddEndPuncttrue
\mciteSetBstMidEndSepPunct{\mcitedefaultmidpunct}
{\mcitedefaultendpunct}{\mcitedefaultseppunct}\relax
\EndOfBibitem
\bibitem[{\v{Z}}ukauskas \latin{et~al.}(2015){\v{Z}}ukauskas,
  Matulaitien{\.{e}}, Paipulas, Niaura, Malinauskas, and Gadonas]{15lpr706}
{\v{Z}}ukauskas,~A.; Matulaitien{\.{e}},~I.; Paipulas,~D.; Niaura,~G.;
  Malinauskas,~M.; Gadonas,~R. Tuning the refractive index in {3D} direct laser
  writing lithography: towards {GRIN} microoptics. \emph{Laser Photon. Rev.}
  \textbf{2015}, \emph{9}, 706--712\relax
\mciteBstWouldAddEndPuncttrue
\mciteSetBstMidEndSepPunct{\mcitedefaultmidpunct}
{\mcitedefaultendpunct}{\mcitedefaultseppunct}\relax
\EndOfBibitem
\bibitem[Nicoletti \latin{et~al.}(2011)Nicoletti, Bulla, Luther-Davies, and
  Gu]{11nl4218}
Nicoletti,~E.; Bulla,~D.; Luther-Davies,~B.; Gu,~M. Generation of $\lambda$/12
  Nanowires in Chalcogenide Glasses. \emph{Nano Lett.} \textbf{2011},
  \emph{11}, 4218--4221\relax
\mciteBstWouldAddEndPuncttrue
\mciteSetBstMidEndSepPunct{\mcitedefaultmidpunct}
{\mcitedefaultendpunct}{\mcitedefaultseppunct}\relax
\EndOfBibitem
\bibitem[Baldacchini \latin{et~al.}(2012)Baldacchini, Snider, and
  Zadoyan]{12oe29890}
Baldacchini,~T.; Snider,~S.; Zadoyan,~R. Two-photon polymerization with
  variable repetition rate bursts of femtosecond laser pulses. \emph{Opt.
  Express} \textbf{2012}, \emph{20}, 29890--29899\relax
\mciteBstWouldAddEndPuncttrue
\mciteSetBstMidEndSepPunct{\mcitedefaultmidpunct}
{\mcitedefaultendpunct}{\mcitedefaultseppunct}\relax
\EndOfBibitem
\bibitem[Mueller \latin{et~al.}(2013)Mueller, Fischer, Mange, Nann, and
  Wegener]{13apl123107}
Mueller,~J.; Fischer,~J.; Mange,~Y.; Nann,~T.; Wegener,~M. In-situ local
  temperature measurement during three-dimensional direct laser writing.
  \emph{Appl. Phys. Lett.} \textbf{2013}, \emph{103}, 123107\relax
\mciteBstWouldAddEndPuncttrue
\mciteSetBstMidEndSepPunct{\mcitedefaultmidpunct}
{\mcitedefaultendpunct}{\mcitedefaultseppunct}\relax
\EndOfBibitem
\bibitem[Sun and Kawata(2004)Sun, and Kawata]{Sun}
Sun,~H.; Kawata,~S. \emph{Two-Photon Photopolymerization and {3D} Lithographic
  Microfabrication}; Springer-Verlag, Berlin, 2004\relax
\mciteBstWouldAddEndPuncttrue
\mciteSetBstMidEndSepPunct{\mcitedefaultmidpunct}
{\mcitedefaultendpunct}{\mcitedefaultseppunct}\relax
\EndOfBibitem
\bibitem[Sun \latin{et~al.}(2003)Sun, Maeda, Takada, Chon, Gu, and
  Kawata]{03apl819}
Sun,~H.-B.; Maeda,~M.; Takada,~K.; Chon,~J.; Gu,~M.; Kawata,~S. Experimental
  investigation of single voxels for laser nanofabrication via two-photon
  photopolymerization. \emph{Appl. Phys. Lett.} \textbf{2003}, \emph{83},
  819\relax
\mciteBstWouldAddEndPuncttrue
\mciteSetBstMidEndSepPunct{\mcitedefaultmidpunct}
{\mcitedefaultendpunct}{\mcitedefaultseppunct}\relax
\EndOfBibitem
\bibitem[Cheng \latin{et~al.}(2014)Cheng, Zeng, Trull, Cojocaru, Malinauskas,
  Jukna, Wiersma, and Staliunas]{14ol6086}
Cheng,~Y.; Zeng,~H.; Trull,~J.; Cojocaru,~C.; Malinauskas,~M.; Jukna,~T.;
  Wiersma,~D.; Staliunas,~K. Beam focalization in reflection from flat
  dielectric subwavelength gratings. \emph{Opt. Lett.} \textbf{2014},
  \emph{39}, 6086--6089\relax
\mciteBstWouldAddEndPuncttrue
\mciteSetBstMidEndSepPunct{\mcitedefaultmidpunct}
{\mcitedefaultendpunct}{\mcitedefaultseppunct}\relax
\EndOfBibitem
\bibitem[Buchegger \latin{et~al.}(2016)Buchegger, Kreutzer, Plochberger,
  Wollhofen, Sivun, Jacak, Schuetz, Schubert, and Klar]{16acsn}
Buchegger,~B.; Kreutzer,~J.; Plochberger,~B.; Wollhofen,~R.; Sivun,~D.;
  Jacak,~J.; Schuetz,~G.; Schubert,~U.; Klar,~T. Stimulated Emission Depletion
  Lithography with Mercapto-Functional Polymers. \emph{{ACS} Nano}
  \textbf{2016}, \emph{10}, 1954–1959\relax
\mciteBstWouldAddEndPuncttrue
\mciteSetBstMidEndSepPunct{\mcitedefaultmidpunct}
{\mcitedefaultendpunct}{\mcitedefaultseppunct}\relax
\EndOfBibitem
\bibitem[Fischer and Wegener(2011)Fischer, and Wegener]{wegener}
Fischer,~J.; Wegener,~M. Three-dimensional direct laser writing inspired by
  stimulated-emission-depletion microscopy. \emph{Opt. Mat. Express}
  \textbf{2011}, \emph{2}, 614--624\relax
\mciteBstWouldAddEndPuncttrue
\mciteSetBstMidEndSepPunct{\mcitedefaultmidpunct}
{\mcitedefaultendpunct}{\mcitedefaultseppunct}\relax
\EndOfBibitem
\bibitem[Wollhofen \latin{et~al.}(2013)Wollhofen, Katzmann, Hrelescu, Jacak,
  and Klar]{13oe10831}
Wollhofen,~R.; Katzmann,~J.; Hrelescu,~C.; Jacak,~J.; Klar,~T. 120 nm
  resolution and 55 nm structure size in {STED}-lithography. \emph{Opt.
  Express} \textbf{2013}, \emph{21}, 10831--10840\relax
\mciteBstWouldAddEndPuncttrue
\mciteSetBstMidEndSepPunct{\mcitedefaultmidpunct}
{\mcitedefaultendpunct}{\mcitedefaultseppunct}\relax
\EndOfBibitem
\bibitem[Xu \latin{et~al.}(2014)Xu, Wang, Ma, Zang, Chen, Lv, Han, Xiao, Zhang,
  Zhang, Ueno, Misawa, and Sun]{Xu}
Xu,~B.~B.; Wang,~L.; Ma,~Z.~C.; Zang,~R.; Chen,~Q.~D.; Lv,~C.; Han,~B.;
  Xiao,~X.~Z.; Zhang,~X.~L.; Zhang,~Y.~L. \latin{et~al.}
  Surface-Plasmon-Mediated Programmable Optical Nanofabrication of an Oriented
  Silver Nanoplate. \emph{ACS Nano} \textbf{2014}, \emph{8}, 6682--6692\relax
\mciteBstWouldAddEndPuncttrue
\mciteSetBstMidEndSepPunct{\mcitedefaultmidpunct}
{\mcitedefaultendpunct}{\mcitedefaultseppunct}\relax
\EndOfBibitem
\bibitem[Xiong \latin{et~al.}(2016)Xiong, Liu, Jiang, Zhou, Li, Jiang, Silvain,
  and Lu]{16am}
Xiong,~W.; Liu,~Y.; Jiang,~L.; Zhou,~Y.; Li,~D.; Jiang,~L.; Silvain,~J.-F.;
  Lu,~Y. Laser-Directed Assembly of Aligned Carbon Nanotubes in Three
  Dimensions for Multifunctional Device Fabrication. \emph{Adv. Mater.}
  \textbf{2016}, 10.1002/adma.201505516\relax
\mciteBstWouldAddEndPuncttrue
\mciteSetBstMidEndSepPunct{\mcitedefaultmidpunct}
{\mcitedefaultendpunct}{\mcitedefaultseppunct}\relax
\EndOfBibitem
\bibitem[\"{O}ktem \latin{et~al.}(2013)\"{O}ktem, Pavlov, Ilday,
  Kalayco\v{g}lu, Rybak, Yavas, Erdo\v{g}an, and Ilday]{Oktem}
\"{O}ktem,~B.; Pavlov,~I.; Ilday,~S.; Kalayco\v{g}lu,~H.; Rybak,~A.; Yavas,~S.;
  Erdo\v{g}an,~M.; Ilday,~F.~O. Nonlinear laser lithography for indefinitely
  large area nanostructuring with femtosecond pulses. \emph{Nature Photon.}
  \textbf{2013}, \emph{7}, 897 -- 901\relax
\mciteBstWouldAddEndPuncttrue
\mciteSetBstMidEndSepPunct{\mcitedefaultmidpunct}
{\mcitedefaultendpunct}{\mcitedefaultseppunct}\relax
\EndOfBibitem
\bibitem[Simanovskii \latin{et~al.}(2003)Simanovskii, Schwettman, Lee, and
  Welch]{Simanovskii}
Simanovskii,~D.; Schwettman,~H.~A.; Lee,~H.; Welch,~A.~J. Midinfrared Optical
  Breakdown in Transparent Dielectrics. \emph{Phys. Rev. Lett.} \textbf{2003},
  \emph{91}, 107601\relax
\mciteBstWouldAddEndPuncttrue
\mciteSetBstMidEndSepPunct{\mcitedefaultmidpunct}
{\mcitedefaultendpunct}{\mcitedefaultseppunct}\relax
\EndOfBibitem
\bibitem[Guay \latin{et~al.}(2012)Guay, Villafranca, Baset, Popov, Ramunno, and
  Bhardwaj]{Guay}
Guay,~J.~M.; Villafranca,~A.; Baset,~F.; Popov,~K.; Ramunno,~L.;
  Bhardwaj,~V.~R. Polarization-dependent femtosecond laser ablation of
  poly-methyl methacrylate. \emph{New J. Phys.} \textbf{2012}, \emph{14},
  085010\relax
\mciteBstWouldAddEndPuncttrue
\mciteSetBstMidEndSepPunct{\mcitedefaultmidpunct}
{\mcitedefaultendpunct}{\mcitedefaultseppunct}\relax
\EndOfBibitem
\bibitem[Buividas \latin{et~al.}(2014)Buividas, Mikutis, and
  Juodkazis]{14pqe119}
Buividas,~R.; Mikutis,~M.; Juodkazis,~S. Surface and bulk structuring of
  materials by ripples with long and short laser pulses: recent advances.
  \emph{Prog. Quant. Electron.} \textbf{2014}, \emph{38}, 119--156\relax
\mciteBstWouldAddEndPuncttrue
\mciteSetBstMidEndSepPunct{\mcitedefaultmidpunct}
{\mcitedefaultendpunct}{\mcitedefaultseppunct}\relax
\EndOfBibitem
\bibitem[Temnov \latin{et~al.}(2006)Temnov, Sokolowski-Tinten, Zhou,
  El-Khamhawy, and von~der Linde]{Temnov}
Temnov,~V.~V.; Sokolowski-Tinten,~K.; Zhou,~P.; El-Khamhawy,~A.; von~der
  Linde,~D. Multiphoton Ionization in Dielectrics: Comparison of Circular and
  Linear Polarization. \emph{Phys. Rev. Lett.} \textbf{2006}, \emph{97},
  237403\relax
\mciteBstWouldAddEndPuncttrue
\mciteSetBstMidEndSepPunct{\mcitedefaultmidpunct}
{\mcitedefaultendpunct}{\mcitedefaultseppunct}\relax
\EndOfBibitem
\bibitem[Gamaly \latin{et~al.}(2010)Gamaly, Juodkazis, Mizeikis, Misawa, Rode,
  and Krolokowski]{10prb054113}
Gamaly,~E.; Juodkazis,~S.; Mizeikis,~V.; Misawa,~H.; Rode,~A.; Krolokowski,~W.
  Modification of refractive index by a single fs-pulse confined inside a bulk
  of a photo-refractive crystal. \emph{Phys. Rev. B} \textbf{2010}, \emph{81},
  054113\relax
\mciteBstWouldAddEndPuncttrue
\mciteSetBstMidEndSepPunct{\mcitedefaultmidpunct}
{\mcitedefaultendpunct}{\mcitedefaultseppunct}\relax
\EndOfBibitem
\bibitem[Gamaly(2011)]{Gam}
Gamaly,~E.~G. The physics of ultra-short laser interaction with solids at
  non-relativistic intensities. \emph{Phys. Reports} \textbf{2011}, \emph{508},
  91 -- 243\relax
\mciteBstWouldAddEndPuncttrue
\mciteSetBstMidEndSepPunct{\mcitedefaultmidpunct}
{\mcitedefaultendpunct}{\mcitedefaultseppunct}\relax
\EndOfBibitem
\bibitem[Gamaly and Rode(2013)Gamaly, and Rode]{Rode}
Gamaly,~E.~G.; Rode,~A.~V. Physics of ultra-short laser interaction with
  matter: From phonon excitation to ultimate transformations. \emph{Progr.
  Quant. Electron.} \textbf{2013}, \emph{37}, 215--323\relax
\mciteBstWouldAddEndPuncttrue
\mciteSetBstMidEndSepPunct{\mcitedefaultmidpunct}
{\mcitedefaultendpunct}{\mcitedefaultseppunct}\relax
\EndOfBibitem
\bibitem[Lifshitz and Pitaevski(1981)Lifshitz, and Pitaevski]{Lifshitz}
Lifshitz,~E.~M.; Pitaevski,~L.~P. \emph{Physical Kinetics}; Pergamon Press,
  Oxford, 1981\relax
\mciteBstWouldAddEndPuncttrue
\mciteSetBstMidEndSepPunct{\mcitedefaultmidpunct}
{\mcitedefaultendpunct}{\mcitedefaultseppunct}\relax
\EndOfBibitem
\bibitem[Mueller \latin{et~al.}(2014)Mueller, Fischer, Mayer, Kadic, and
  Wegener]{14am6566}
Mueller,~J.; Fischer,~J.; Mayer,~F.; Kadic,~M.; Wegener,~M. Polymerization
  Kinetics in Three-Dimensional Direct Laser Writing. \emph{Adv. Mater.}
  \textbf{2014}, \emph{26}, 6566--6571\relax
\mciteBstWouldAddEndPuncttrue
\mciteSetBstMidEndSepPunct{\mcitedefaultmidpunct}
{\mcitedefaultendpunct}{\mcitedefaultseppunct}\relax
\EndOfBibitem
\bibitem[Malinauskas \latin{et~al.}(2016)Malinauskas, \v{Z}ukauskas, Hasegawa,
  Hayasaki, Mizeikis, Buividas, and Juodkazis]{16lsa}
Malinauskas,~M.; \v{Z}ukauskas,~A.; Hasegawa,~S.; Hayasaki,~Y.; Mizeikis,~V.;
  Buividas,~R.; Juodkazis,~S. Ultrafast laser processing of materials: from
  science to industry. \emph{Light: Sci. Appl.} \textbf{2016}, \emph{5},
  e16133\relax
\mciteBstWouldAddEndPuncttrue
\mciteSetBstMidEndSepPunct{\mcitedefaultmidpunct}
{\mcitedefaultendpunct}{\mcitedefaultseppunct}\relax
\EndOfBibitem
\bibitem[Nasse and Woehl(2010)Nasse, and Woehl]{Nasse10}
Nasse,~M.~J.; Woehl,~J.~C. Realistic modeling of the illumination point spread
  function in confocal scanning optical microscopy. \emph{J. Opt. Soc. Am. A}
  \textbf{2010}, \emph{27}, 295\relax
\mciteBstWouldAddEndPuncttrue
\mciteSetBstMidEndSepPunct{\mcitedefaultmidpunct}
{\mcitedefaultendpunct}{\mcitedefaultseppunct}\relax
\EndOfBibitem
\bibitem[Stuart. \latin{et~al.}(1995)Stuart., Feit, Rubenchick, Shore, and
  Perry]{Stuart}
Stuart.,~B.~C.; Feit,~M.~D.; Rubenchick,~A.~M.; Shore,~B.~W.; Perry,~M.~D.
  Laser-induced damage in dielectrics with nanosecond to picosecond pulses.
  \emph{Phys. Rev. Lett.} \textbf{1995}, \emph{74}, 2248--2251\relax
\mciteBstWouldAddEndPuncttrue
\mciteSetBstMidEndSepPunct{\mcitedefaultmidpunct}
{\mcitedefaultendpunct}{\mcitedefaultseppunct}\relax
\EndOfBibitem
\bibitem[Luther-Davies \latin{et~al.}(2005)Luther-Davies, Rode, Madsen, and
  Gamaly]{Gamaly}
Luther-Davies,~B.; Rode,~A.; Madsen,~N.; Gamaly,~E.~G. Picosecond
  high-repitition-rate pulsed laser ablation of dielectrics: the effect of
  energy accumulation between pulses. \emph{Opt. Eng.} \textbf{2005},
  \emph{44}, 051102\relax
\mciteBstWouldAddEndPuncttrue
\mciteSetBstMidEndSepPunct{\mcitedefaultmidpunct}
{\mcitedefaultendpunct}{\mcitedefaultseppunct}\relax
\EndOfBibitem
\bibitem[Ovsianikov \latin{et~al.}(2009)Ovsianikov, Shizhou, Farsari,
  Vamvakaki, Fotakis, and Chichkov]{09oe2143}
Ovsianikov,~A.; Shizhou,~X.; Farsari,~M.; Vamvakaki,~M.; Fotakis,~C.;
  Chichkov,~B. Shrinkage of microstructures produced by two-photon
  polymerization of Zr-based hybrid photosensitive materials. \emph{Opt.
  Express} \textbf{2009}, \emph{17}, 2143--2148\relax
\mciteBstWouldAddEndPuncttrue
\mciteSetBstMidEndSepPunct{\mcitedefaultmidpunct}
{\mcitedefaultendpunct}{\mcitedefaultseppunct}\relax
\EndOfBibitem
\bibitem[Chon \latin{et~al.}(2002)Chon, Gan, and Gu]{Chon}
Chon,~J. W.~M.; Gan,~X.; Gu,~M. Splitting of the focal spot of a high
  numerical-aperture objective in free space. \emph{Appl. Phys. Lett.}
  \textbf{2002}, \emph{81}, 1576 -- 1578\relax
\mciteBstWouldAddEndPuncttrue
\mciteSetBstMidEndSepPunct{\mcitedefaultmidpunct}
{\mcitedefaultendpunct}{\mcitedefaultseppunct}\relax
\EndOfBibitem
\bibitem[Ze'ev \latin{et~al.}(2006)Ze'ev, Gu, and Shamir]{Shamir}
Ze'ev,~B.; Gu,~M.; Shamir,~J. Angular momentum and geometrical phases in
  tight-focused circularly polarized plane waves. \emph{Appl. Phys. Lett.}
  \textbf{2006}, \emph{89}, 241104\relax
\mciteBstWouldAddEndPuncttrue
\mciteSetBstMidEndSepPunct{\mcitedefaultmidpunct}
{\mcitedefaultendpunct}{\mcitedefaultseppunct}\relax
\EndOfBibitem
\bibitem[Gibbon(2005)]{Gibbon}
Gibbon,~P. \emph{Short pulse laser interactions with Matter}; Imperial College
  Press, London, 2005\relax
\mciteBstWouldAddEndPuncttrue
\mciteSetBstMidEndSepPunct{\mcitedefaultmidpunct}
{\mcitedefaultendpunct}{\mcitedefaultseppunct}\relax
\EndOfBibitem
\bibitem[Jia \latin{et~al.}(2005)Jia, Chen, Huang, Zhao, Qiu, Li, Xu, He,
  Zhang, and Kuroda]{05prb125429}
Jia,~T.; Chen,~H.; Huang,~M.; Zhao,~F.; Qiu,~J.; Li,~R.; Xu,~Z.; He,~X.;
  Zhang,~J.; Kuroda,~H. Formation of nanogratings on the surface of a {ZnSe}
  crystal irradiated by femtosecond laser pulses. \emph{Phys. Rev. B}
  \textbf{2005}, \emph{72}, 125429\relax
\mciteBstWouldAddEndPuncttrue
\mciteSetBstMidEndSepPunct{\mcitedefaultmidpunct}
{\mcitedefaultendpunct}{\mcitedefaultseppunct}\relax
\EndOfBibitem
\bibitem[Lee \latin{et~al.}(2007)Lee, Yang, and Nikumb]{07ass4740}
Lee,~S.; Yang,~D.; Nikumb,~S. Femtosecond laser patterning of
  ${T}a_{0.1}{W}_{0.9}{O}_{x}$/{ITO} thin film stack. \emph{Appl. Surf. Sci.}
  \textbf{2007}, \emph{253}, 4740--4747\relax
\mciteBstWouldAddEndPuncttrue
\mciteSetBstMidEndSepPunct{\mcitedefaultmidpunct}
{\mcitedefaultendpunct}{\mcitedefaultseppunct}\relax
\EndOfBibitem
\bibitem[Graf and Muller(2015)Graf, and Muller]{15ass}
Graf,~S.; Muller,~F. Polarization-dependent generation of fs-laser induced
  periodic surface structures. \emph{Appl. Surf. Sci.} \textbf{2015},
  \emph{331}, 150--155\relax
\mciteBstWouldAddEndPuncttrue
\mciteSetBstMidEndSepPunct{\mcitedefaultmidpunct}
{\mcitedefaultendpunct}{\mcitedefaultseppunct}\relax
\EndOfBibitem
\bibitem[Maigyte \latin{et~al.}(2013)Maigyte, Purlys, Trull, Peckus, Cojocaru,
  Gailevi{\v{c}}ius, Malinauskas, and Staliunas]{13ol2376}
Maigyte,~L.; Purlys,~V.; Trull,~J.; Peckus,~M.; Cojocaru,~C.;
  Gailevi{\v{c}}ius,~D.; Malinauskas,~M.; Staliunas,~K. Flat lensing in the
  visible frequency range by woodpile photonic crystals. \emph{Opt. Lett.}
  \textbf{2013}, \emph{38}, 2376--2378\relax
\mciteBstWouldAddEndPuncttrue
\mciteSetBstMidEndSepPunct{\mcitedefaultmidpunct}
{\mcitedefaultendpunct}{\mcitedefaultseppunct}\relax
\EndOfBibitem
\bibitem[Ovsianikov \latin{et~al.}(2008)Ovsianikov, Viertl, Chichkov, Oubaha,
  MacCraith, Sakellari, Giakoumaki, Gray, Vamvakaki, Farsari, and
  Fotakis]{08acsn2257}
Ovsianikov,~A.; Viertl,~J.; Chichkov,~B.; Oubaha,~M.; MacCraith,~B.;
  Sakellari,~I.; Giakoumaki,~A.; Gray,~D.; Vamvakaki,~M.; Farsari,~M.
  \latin{et~al.}  Ultra-Low Shrinkage Hybrid Photosensitive Material for
  Two-Photon Polymerization Microfabrication. \emph{{ACS Nano}} \textbf{2008},
  \emph{2}, 2257--2262\relax
\mciteBstWouldAddEndPuncttrue
\mciteSetBstMidEndSepPunct{\mcitedefaultmidpunct}
{\mcitedefaultendpunct}{\mcitedefaultseppunct}\relax
\EndOfBibitem
\end{mcitethebibliography}

\end{document}